\documentclass{article}
\usepackage{spconf,amsmath,graphicx}
\usepackage{url}
\usepackage{multirow}
\usepackage{svg}
\usepackage{amsmath}
\usepackage[list-final-separator={, and }]{siunitx}
\usepackage{float}
\usepackage{tikz}
\usepackage{subfig}
\usepackage{todonotes}
\usepackage{bm}

\DeclareSIUnit{\pp}{\textup{p.p.}}

\title{Deep Learning for High Speed Optical Coherence Elastography}

\name{M. Neidhardt*, M. Bengs*, S. Latus, M. Schl\"uter, T. Saathoff, and A. Schlaefer \thanks{* Both authors contributed equally.}}
\address{Institute of Medical Technology, Hamburg University of Technology, Germany}
\begin{document}
%
\maketitle
\begin{abstract}
Mechanical properties of tissue provide valuable information for identifying lesions. One approach to obtain quantitative estimates of elastic properties is shear wave elastography with optical coherence elastography (OCE). However, given the shear wave velocity, it is still difficult to estimate elastic properties. Hence, we propose deep learning to directly predict elastic tissue properties from OCE data. We acquire 2D images with a frame rate of 30 kHz and use convolutional neural networks to predict gelatin concentration, which we use as a surrogate for tissue elasticity. We compare our deep learning approach to predictions from conventional regression models, using the shear wave velocity as a feature. Mean absolut prediction errors for the conventional approaches range from \SI[separate-uncertainty = true,multi-part-units=single]{1,32(98)}{\pp} to \SI[separate-uncertainty = true,multi-part-units=single]{1,57(130)}{\pp} whereas we report an error of \SI[separate-uncertainty = true,multi-part-units=single]{0,90(84)}{\pp} for the convolutional neural network with 3D spatio-temporal input. Our results indicate that deep learning on spatio-temporal data outperforms elastography based on explicit shear wave velocity estimation.

\end{abstract}

\section{Introduction}
\label{sec:intro}

Diseases as steatosis or cancer cause tissue degeneration and thereby change the mechanical properties of tissue. Therefore, information about these properties can be used for instance in discriminating different fibrosis stages \cite{franchi2015}. Elastography methods allow the measurement of tissue stiffness. One approach is based on compression of tissue and estimating relative tissue elasticities. Nevertheless, a quantitative estimation of elastic properties is not possible with this approach. Another approach is shear wave elastography where the shear wave velocity allows for a quantitative estimate of tissue elasticity. 

For imaging propagating shear waves we consider optical coherence tomography (OCT) as imaging modality which provides a high temporal and spatial resolution. Shear wave velocities can be estimated from OCT phase data with conventional image processing methods. One approach is to determine the time of flight between wave excitation and displacement at distinct measurement points \cite{latus2017}. Other studies obtain the velocity by tracking the wavefront of a shear wave across an OCT B-Scan with high temporal resolution \cite{song2013shear}. Several models have been proposed that map explicit features such as the shear wave velocity to elastic properties, e.g., Young's modulus \cite{han2016}. 

 However, shear wave velocity estimation is challenging due to the distance to the surface of the wave, tissue inhomogeneities, jitter, and other motion. Moreover, material stiffness estimation is often restricted to the strong theoretical assumption of an isotropic homogeneous Voigt material model and assumes a negligible viscosity of soft tissue \cite{wang2015optical}. Further, potentially valuable information such as reverberations, wave amplitudes, and lengths as well as wave dependencies along the depth dimension are neglected. Therefore, we propose to use deep learning in order to avoid explicit feature extraction and complex image processing. Deep learning methods have been studied in the field of ultrasound elastography \cite{pereira2018}. Nevertheless, so far in OCE only simple classification methods based on feature extraction and OCT image data have been demonstrated \cite{liu2016}. We are the first to present deep learning on OCE image data without feature extraction to predict elastic tissue properties.
 
 In this work, we consider 2D and 3D OCT image sequences combined with deep learning to directly estimate material properties. We evaluate our methods on gelatin phantoms with varying stiffness with an experimental setup for OCT elastography (OCE). We predict gelatin concentration in different samples, which serves as a surrogate for the elastic properties and samples can be created easily and reproducibly. We compare deep learning approaches to predictions from linear regression, support vector regression, and multi layer perceptrons using the shear wave velocity as explicit feature. We illustrate that deep learning substantially improves the performance and simplifies the overall process of OCE with respect to the signal processing. 

\section{Material and Methods}
\label{sec:format}
\subsection{Data and feature extraction} 
We employ OCT for imaging a propagating wave and define an A-scan as a one dimensional depth resolved signal. By moving the OCT light beam along a line and acquiring a sequence of A-scans, we obtain a two dimensional image (B-scan). We acquire B-scans at a scan rate of \SI{30}{kHz} and refer to this data structure as 2D+$t$ in the following sections. All methods are only evaluated with the phase part of the OCT signal, referred to as OCE phase data in the following. We unwrap phase data along the time axis and calculate the phase difference between subsequent B-scans for each pixel. 

\begin{figure}[tb!]
\centering
\includegraphics[width=.95\linewidth]{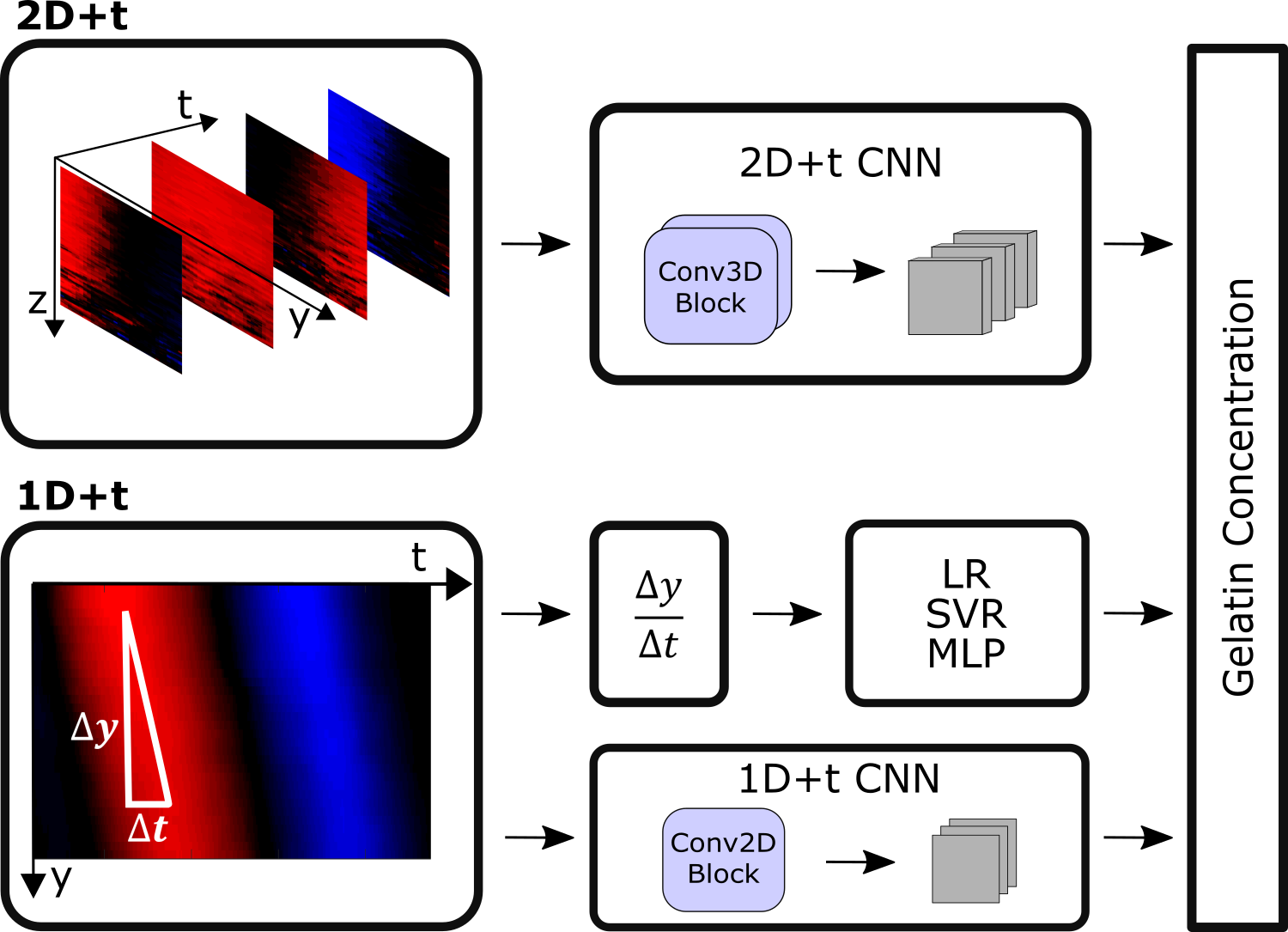}
\caption{Proposed methods for mapping of OCE 2D+$t$ and OCE 1D+$t$ data to gelatin concentration.}
\label{fig:2D_3D_Data}
\end{figure}

In addition to the image data, we consider shear wave velocity as an explicit feature for prediction. Extraction of the shear wave velocity is performed similar to Nguyen et al.\ \cite{nguyen2015}. To improve the overall signal-to-noise ratio we eliminate rows by thresholding in the B-scans which have a low signal quality due to poor speckle. Remaining noise is reduced by applying a 3D median filter of kernel size \num{3} along all directions. Exemplary B-scans are shown in Fig.~\ref{fig:2D_3D_Data} (top left). Next, we compute the mean along the axial $z$-direction which gives us a 1D+$t$ representation with a size of $32\!\times\!1\!\times\!400$ pixels along the $y$, $z$ and $t$ direction, respectively. The resulting 1D+$t$ spatio-temporal representation can be seen in Fig.~\ref{fig:2D_3D_Data} (bottom left). Next, we detect the peaks of the initial wavefront and estimate the shear wave velocity $v = \Delta y / \Delta t$ with a linear regression approach. Consequently, we extract the shear wave velocity as the number of pixels the first wave front travels between two adjacent B-scans over time.

\subsection{Deep learning architectures and regression methods}
We evaluate and compare two deep learning models for predicting the gelatin concentration of phantoms based on 1D+$t$ or 2D+$t$ phase data. We consider our learning task as a regression problem of gelatin concentration.

We use a state-of-the art architecture as our baseline and choose the idea of densely connected neural networks \cite{Huang2017}. Our architecture consists of one initial convolutional layer with a kernel size of five and three for the temporal and the spatial dimensions, respectively. For the temporal dimension, we use a kernel stride of four in the first layer for downsampling. The initial convolutional layer is followed by four DenseNet-blocks with a growth rate of 5 and four layers each. We connect the DenseNet-blocks with average pooling layers. We use a global average pooling layer (GAP) prior to the regression layer with one scalar output. For this baseline architecture, we evaluate 2D and 3D spatio-temporal convolutions for 1D+$t$ and 2D+$t$ phase data, respectively. In particular, employing spatio-temporal convolutions enables to learn features jointly from the spatial and temporal dimension. We refer to the 1D+$t$ model as 1D+$t$ CNN and to the 2D+$t$ as 2D+$t$ CNN.

Moreover, we employ three different conventional regression methods for predicting gelatin concentration based on shear wave velocity. First, we employ a linear regression (LR). Second, we use support vector regression (SVR) and evaluate a linear kernel and a Gaussian kernel (RBF). Third, we use multilayer perceptron neural network (MLP) regression with two hidden layers. We evaluate two networks with 50 and 100 hidden units. For all our neural networks we use Adam for optimization with a batch size of ten. 

\subsection{Experimental setup and data acquisition}
\begin{figure*}[t!]
\includegraphics[width=.39\linewidth]{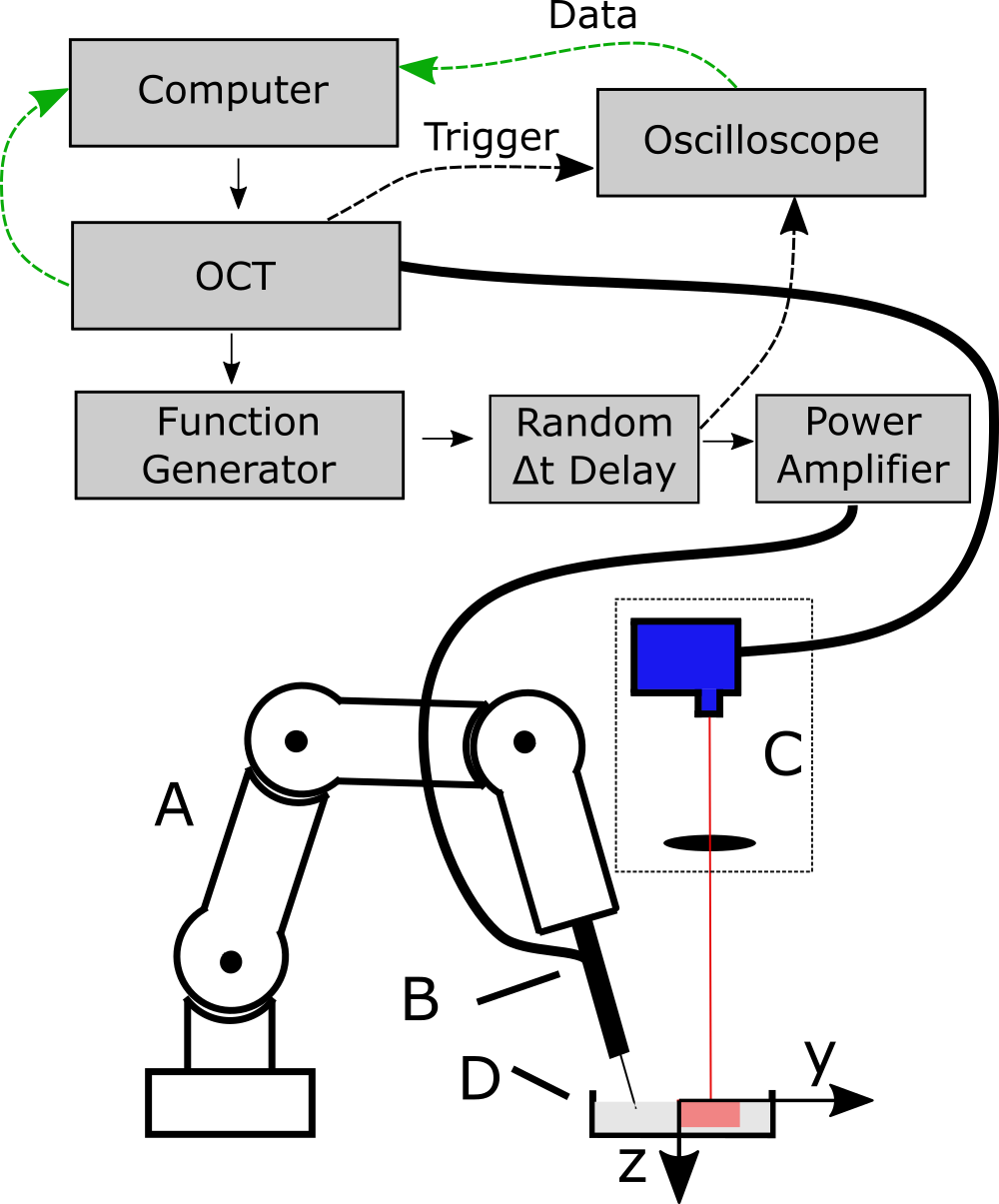}
\includegraphics[width=.59\linewidth]{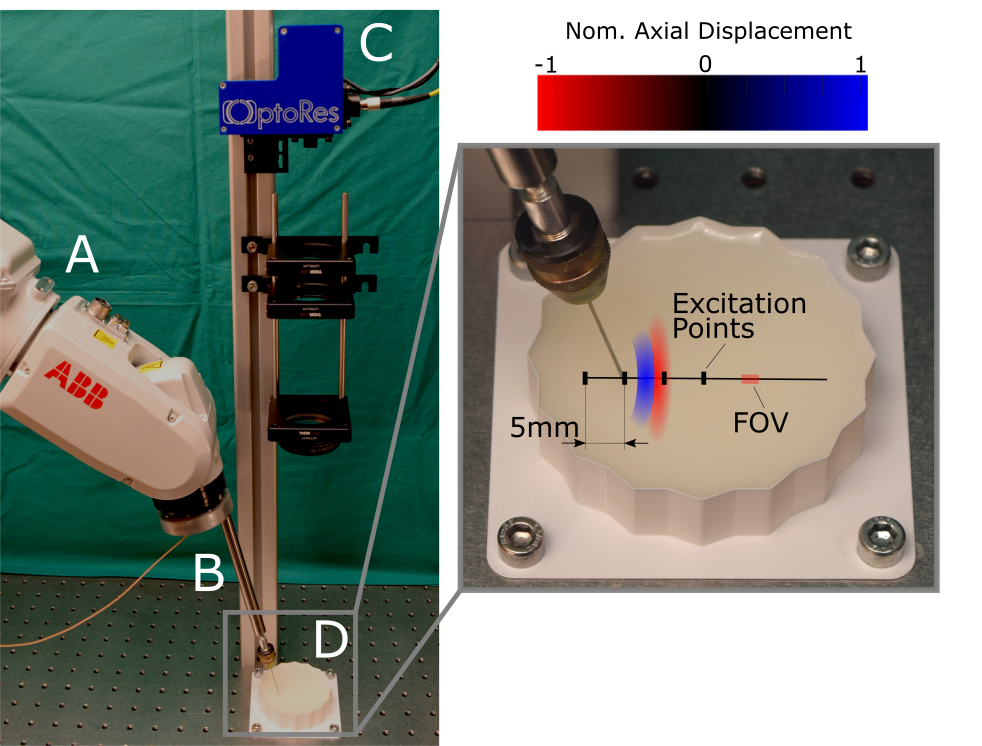}
\centering
\caption{Experimental setup for data acquisition. Top left: The \SI{1.59}{MHz} OCT system triggers a piezoelectric actuator and an oscilloscope records the trigger signals of all devices. Bottom left: Schematic setup with (A) robot, (B) piezoelectric actuator and needle, (C) OCT scan head, and (D) gelatin sample. The red square marks the OCT's FOV with the lateral scan direction indicated as $y$.
Middle: Picture of the experimental setup. Right: The needle is positioned at four excitation points. The excited shear wave, indicated with a blue and a red stripe, propagates cylindrically from left to right through the FOV.}
 \label{fig:Setup_Drawing}
\end{figure*}

For data acquisition we propose the setup depicted in Fig.~\ref{fig:Setup_Drawing}. Imaging is performed with a high-speed swept-source OCT device (OMES, Optores GmbH) with an A-scan rate of \SI{1.59}{MHz} and a central wavelength of \SI{1315}{\nano\meter}. A scan head acquired B-scans with an effective size of $32\!\times\!250\!\times\!1$ pixels along the $y$, $z$ and $t$ direction, respectively, and a field of view (FOV) of approximately \SI{3x2}{mm} in air. We use a lens with a focal length of \SI{300}{mm} for focusing the beams. 

Shear waves are generated with a clinical needle (gauge 21) attached to a piezoelectric actuator. We mount the actuator on a robot arm for automatic and reproducible placement of the needle. A single OCE measurement consists of the following steps. First, the robot arm inserts the needle approximately \SI{3}{mm} within a gelatin phantom. Second, the OCT system is set to acquire \num{200000} A-scans and transmits a start trigger to a function generator (RedPitaya, StemLabs) to generate a single burst function for shear wave excitation. The burst signal is amplified resulting in a movement of about \SI{50}{\micro\meter} along the needle shaft with an actuator modulation frequency of \SI{100}{Hz}. An oscilloscope records the excitation burst and B-scan acquisition trigger of the OCT device to synchronize data consistently.

For data generation we prepare tissue mimicking phantoms with a gelatin to water ratio of 1:8 (11.1\%), 1:11 (8.3\%), 1:14 (6.7\%), 1:17 (5.6\%), 1:20 (4.8\%), and 1:23 (4.2\%). The number in brackets refers to the gelatin concentration in percent. We add TiO$_{2}$ particles to increase speckle in the OCE data. Moreover, we use two samples per gelatin concentration and evaluate each sample with two different orientations w.r.t.\ the OCT's FOV. Using the robot, we evaluate four different needle positions: \SIlist{5;10;15;20}{\milli\meter} away from the FOV (Fig.~\ref{fig:Setup_Drawing}, right). We acquire four measurements for each combination of needle position and sample orientation. Hence, we perform 64 measurements in total for each concentration. Note that a random time delay between OCT acquisition and shear wave excitation allows us to acquire OCE data of the travelling wave at various wave cycles within the FOV and further increases the variation in our data. 

To reduce computational effort, the following data preprocessing steps are applied. We crop each 2D+$t$ OCE data to \num{400} B-scans corresponding to \SI{13.3}{ms}. Furthermore, we crop B-Scans to only contain gelatin speckle information by removing image data above the phantom surface. This step also prevents unintended correlation between the position of a phantom's surface in the B-scan and its concentration.

\subsection{Training and evaluation}
To test our different methods on previously unseen gelatin concentrations, we apply a six-fold cross validation approach. For each fold we leave out the data for one gelatin concentration for testing and validating and use the data for the remaining five concentrations for optimization. We split the data equally into a test and validation subset for each fold, hence each subsets consists of 32 samples. We leverage the validation subset for hyperparameter tuning of our methods. 

\section{Results}
\label{sec:pagestyle}
\begin{figure}[t!]
\includegraphics[width=.95\linewidth]{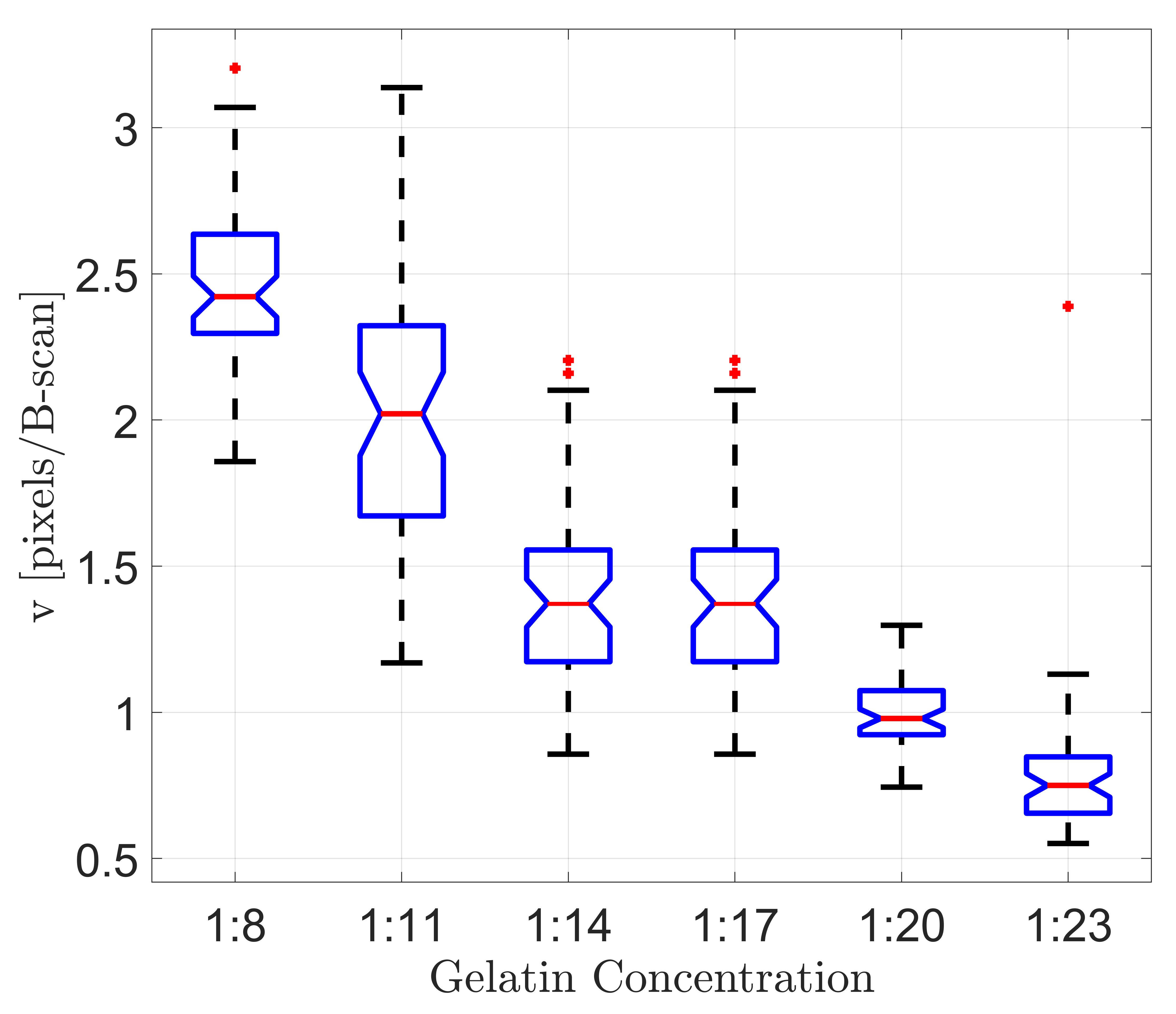}
\centering
\caption{Shear wave velocity estimated by explicit feature extraction. The gelatin concentration is given as the ratio of gelatin to water.}
 \label{fig:velocities}
\end{figure}

Results of our explicit feature extraction approach are shown in Fig.~\ref{fig:velocities}, which yields a decreasing shear wave velocity for softer phantoms with a lower gelatin concentrations. Further, our predictions in gelatin concentration are reported in Table~\ref{tab:All-networks-with metrics} by the mean absolute error (MAE), relative mean absolute error (rMAE), and average correlation coefficient (ACC) for all experiments and applied methods. Note that we average the metrics over all six test folds. The MAE is given in percentage points of the gelatin concentration. The rMAE is relative to the target's standard deviation. Consistently over all metrics, the 2D+$t$ spatio-temporal CNN performs best. The MLP regression shows the best performance of the different regression methods employing the estimated shear wave velocities. 

\begin{table}[t!]
\centering
\caption{Results for all experiments. The MAE refers to predicted gelatin concentration and given in percentage points (p.p.).} 
\label{tab:All-networks-with metrics}%
\begin{tabular}{lccc}
  & MAE (p.p.)  & rMAE & ACC  \\ \hline
  LR & $1.57\pm1.30$ &$ 0.67\pm0.55$ & $0.57$  \\
  SVR (Linear) & $1.50\pm1.30$ &$ 0.63\pm0.54$ & $0.60$  \\ 
  SVR (RBF) & $1.41\pm1.10$ &$ 0.60\pm0.46$ & $0.68$  \\
  MLP (50,50) & $1.29\pm1.02$ &$ 0.55\pm0.43$ & $0.73$  \\
  MLP (100,100) & $1.32\pm0.98$ &$ 0.60\pm0.41$ & $0.73$  \\
  \hline
  1D+$t$ CNN & $1.04\pm0.86$ &$ 0.44\pm0.37$ & $0.83$  \\
   \textbf{2D+$\bm{t}$ CNN} & \bm{$0.90\pm0.84$} & \bm{$ 0.38\pm0.36$} &  \bm{$0.87$}  \\
\hline
  \end{tabular}
\end{table}

\section{Discussion and Conclusion}
\label{sec:conclusion}
We consider the task to estimate material properties from OCE data with conventional regression methods from shear wave velocities and a deep learning approach on image data. Fig.~\ref{fig:velocities} highlights the dependency of shear wave velocity on gelatin concentration, as already demonstrated in previous studies \cite{latus2017,nguyen2015}. However, to differentiate between estimated shear wave velocities between 5.6$\%$ and 6.7$\%$ of gelatin concentration already seems to be difficult. In this regard, the evaluation of the different conventional regression methods indicates that estimating and using shear wave velocity for tissue characterization is a challenging task. While our non-linear MLP approach achieves the best performance of our conventional regression methods, our linear methods do not perform well. This suggests a complex and non-linear relationship between gelatin concentrations and estimated shear wave velocities, influenced by measurement noise and possible inhomogeneities. Moreover, the performance is increased with more complex deep learning models and OCE image data. In this regard, a 1D+$t$ image representation combined with a 1D+$t$ CNN already improves the performance notably and unmodified 2D+$t$ OCE phase data with a 2D+$t$ spatio-temporal CNN performs best. The performance difference between our 2D+$t$ and 3D+$t$ CNN indicates, that pre-processing from 2D+$t$ to 1D+$t$ removes information.

Our results illustrate that material characterization by shear wave velocity is challenging and that deep learning using 2D+$t$ OCE phase data directly presents a promising and simpler alternative to conventional signal processing.


\bibliographystyle{IEEEbib}
\bibliography{strings,refs}

\begin{thebibliography}{1}

\bibitem{franchi2015}
S.~Franchi-Abella, L.~Corno, E.~Gonzales, G.~Antoni, M.~Fabre, B.~Ducot,
  D.~Pariente, J.L. Gennisson, M.~Tanter, and J.M. Corr{\'e}as,
\newblock ``Feasibility and diagnostic accuracy of supersonic shear-wave
  elastography for the assessment of liver stiffness and liver fibrosis in
  children: a pilot study of 96 patients,''
\newblock {\em Radiology}, vol. 278, no. 2, pp. 554--562, 2015.

\bibitem{latus2017}
S.~Latus, C.~Otte, M.~Schl{\"u}ter, J.~Rehra, K.~Bizon, H.~Schulz-Hildebrandt,
  T.~Saathoff, G.~H{\"u}ttmann, and A.~Schlaefer,
\newblock ``An approach for needle based optical coherence elastography
  measurements,''
\newblock in {\em International Conference on Medical Image Computing and
  Computer-Assisted Intervention}. Springer, 2017, pp. 655--663.

\bibitem{song2013shear}
S.~Song, Z.~Huang, T.-M. Nguyen, E.Y. Wong, B.~Arnal, M.~O'Donnell, and R.K.
  Wang,
\newblock ``Shear modulus imaging by direct visualization of propagating shear
  waves with phase-sensitive optical coherence tomography,''
\newblock {\em Journal of Biomedical Optics}, vol. 18, no. 12, pp. 121509,
  2013.

\bibitem{han2016}
Z.~Han, M.~Singh, S.~R. Aglyamov, C.~Liu, A.~Nair, R.~Raghunathan, C.~Wu,
  J.~Li, and K.~V. Larin,
\newblock ``Quantifying tissue viscoelasticity using optical coherence
  elastography and the rayleigh wave model,''
\newblock {\em Journal of Biomedical Optics}, vol. 21, no. 9, pp. 090504, 2016.

\bibitem{wang2015optical}
S.~Wang and K.V. Larin,
\newblock ``Optical coherence elastography for tissue characterization: A
  review,''
\newblock {\em Journal of Biophotonics}, vol. 8, no. 4, pp. 279--302, 2015.

\bibitem{pereira2018}
C.~Pereira, M.~Dighe, and A.M. Alessio,
\newblock ``Comparison of machine learned approaches for thyroid nodule
  characterization from shear wave elastography images,''
\newblock in {\em Medical Imaging 2018: Computer-Aided Diagnosis}.
  International Society for Optics and Photonics, 2018, vol. 10575, p. 105751X.

\bibitem{liu2016}
C.~Liu, Y.~Du, M.~Singh, C.~Wu, Z.~Han, J.~Li, A.~Chang, C.~Mohan, and K.V.
  Larin,
\newblock ``Classifying murine glomerulonephritis using optical coherence
  tomography and optical coherence elastography,''
\newblock {\em Journal of biophotonics}, vol. 9, no. 8, pp. 781--791, 2016.

\bibitem{nguyen2015}
T.~Nguyen, B.~Arnal, S.~Song, Z.~Huang, R.K. Wang, and M.~O’Donnell,
\newblock ``Shear wave elastography using amplitude-modulated acoustic
  radiation force and phase-sensitive optical coherence tomography,''
\newblock {\em Journal of Biomedical Optics}, vol. 20, no. 1, pp. 016001, 2015.

\bibitem{Huang2017}
G.~Huang, Z.~Liu, L.~Van Der~Maaten, and K.Q. Weinberger,
\newblock ``{Densely Connected Convolutional Networks},''
\newblock {\em Conference on Computer Vision and Pattern Recognition}, pp.
  2261--2269, 2017.

\end{thebibliography}

\pagebreak

\end{document}